\documentclass[floatfix,prb,twocolumn,showpacs]{revtex4-1}

\usepackage{amsmath}
\usepackage{amssymb}
\usepackage{graphicx}
\usepackage{dcolumn}
\usepackage{natbib}

\newcommand{\ee}{\mathrm{e}}
\newcommand{\dd}{\mathrm{d}}
\newcommand{\ii}{\mathrm{i}}
\newcommand{\bx}{{\boldsymbol x}}
\newcommand{\by}{{\boldsymbol y}}

\begin{document}

\title{Comprehensive study of the critical behavior in the diluted antiferromagnet in a field}

\author{L.~A.~Fernandez} 
\author{V.~Martin-Mayor} 
\author{D.~Yllanes}
  \affiliation{Departamento de F\'\i{}sica Te\'orica I, Universidad
  Complutense, 28040 Madrid, Spain.}
  \affiliation{Instituto de Biocomputaci\'on y F\'{\i}sica de Sistemas
  Complejos (BIFI), 50018 Zaragoza, Spain.}

\begin{abstract}
We study the critical behavior of the Diluted Antiferromagnet in a Field
with the Tethered Monte Carlo formalism. 
We compute the critical exponents (including the elusive hyperscaling 
violations exponent $\theta$).  Our results provide a comprehensive 
description of the phase transition and clarify the inconsistencies 
between previous experimental and theoretical work. To do so, our method
addresses the usual problems of numerical work (large tunneling barriers and self-averaging violations).

\end{abstract}


\maketitle

Understanding collective behavior in the presence of quenched disorder has
long been one of the most challenging and interesting problems in statistical
mechanics. One of its simplest representatives is the random field Ising model
(RFIM), which has been extensively studied both theoretically and
experimentally.\cite{nattermann:97,*belanger:97} The RFIM is physically
realized by a diluted antiferromagnet in an applied magnetic field (DAFF).

It is known that the $D=3$ DAFF/RFIM undergoes a phase transition,
but the details remain controversial, with severe inconsistencies between analytical, 
experimental and numerical work. A scaling theory is generally accepted, where 
the dimension $D$ of the system is replaced by $D-\theta$ in the hyperscaling 
relation. This third independent critical exponent, 
believed to be $\theta\approx1.5$,
is inaccessible both to a direct experimental measurement and to traditional
Monte Carlo methods.

The values of the remaining critical exponents, seemingly  more 
straightforward, are also controversial. On the experimental front,
different ans\"atze for the scattering line shape yield
mutually incompatible estimates of the thermal critical exponent,
namely $\nu=0.87(7)$ (Ref.~\onlinecite{slanic:99}), or $\nu=1.20(5)$ (Ref.~\onlinecite{ye:04}).
Furthermore, the experimental estimate of the anomalous dimension,
$\eta=0.16(6)$ (Ref.~\onlinecite{slanic:99}), violates hyperscaling bounds,
if one is to believe the experimental claims of a diverging 
specific heat ($\alpha\geq0$).~\cite{belanger:83,*belanger:98}

On the other hand, the numerical determination of $\nu$ has steadily shifted,
the most precise estimate being $1.37(9)$ (Ref.~\onlinecite{middleton:02}),
inconsistent with the experimental values and barely compatible
with $\alpha\approx0$. The value of $\alpha$ itself
is very hard to measure in a numerical
simulation.~\cite{hartmann:01,*malakis:06}

More fundamentally, the smallness of the magnetic exponent $\beta$, combined
with the numerical observation of
metastability,~\cite{sourlas:99,*wu:06,*maiorano:07} has led some authors to
suggest that the transition in the DAFF may be of first order.

Ultimately, the physical reasons for this confusion betray the fact that the
traditional tools of statistical mechanics are ill-suited to systems with
rugged free-energy landscapes.  Both experimentally and numerically, the
system gets trapped in local minima, with escape times that grow as $\log \tau
\sim \xi^\theta$ ($\xi$ is the correlation length). This not only makes it
exceedingly hard to thermalize the system, but also generates a rare-events
statistics, causing self-averaging violations.~\cite{parisi:02,*fytas:11}

In this letter we study the DAFF with
the Tethered Monte Carlo (TMC) formalism.~\cite{fernandez:09}
Our approach restores self-averaging and is able to negotiate
the free-energy barriers of the DAFF to 
equilibrate large systems safely. It also provides direct access to the
key parameter $\theta$. We thus obtain a comprehensive 
picture of the phase transition, consistent both with analytical results 
for the RFIM and with experiments on the DAFF, and shed 
light on the reasons behind the previous discrepancies.

In the following we provide a brief outline of the tethered formalism applied
to the DAFF (see Refs.~\onlinecite{fernandez:09,martin-mayor:11} for details).  We note,
however, that we give most of our physical results translated into the familiar
canonical language.  In a tethered computation, we run simulations where
one (or more) order parameters of the system are (almost) constrained. In this
way, we eliminate the need for exponentially slow tunneling caused by the
free-energy barriers associated to these parameters. From these tethered
simulations the Helmholtz effective potential is accurately reconstructed with
a fluctuation-dissipation formalism.

We  consider a system with $N=L^D$ spins, $s_\bx=\pm1$, on the nodes
of a cubic lattice with periodic boundary conditions and interacting
through the Hamiltonian
\begin{equation}\label{eq:H}
H = \sum_{\langle \bx,\by\rangle} \epsilon_\bx s_\bx \epsilon_\by s_\by- h M - h_\text{s} M_\text{s}
= U - h M - h_\text{s} M_\text{s}.
\end{equation}
Here $h$ and $h_\text{s}$ are the applied fields, coupled to the magnetization and staggered magnetization,
\begin{align}
M &= Nm= \sum_\bx \epsilon_\bx s_\bx, &
M_\text{s} &= \sum_\bx  \epsilon_\bx s_\bx \ee^{\ii\pi \sum_{\mu=1}^D x_\mu}.
\end{align}
We are ultimately interested in $h_\text{s}=0$, but we will find
this parameter useful.  The quenched occupation variables $\epsilon_\bx$ 
are $1$ with probability $p=0.7$ and zero otherwise
(this value is chosen to be far 
both from the percolation threshold and from the pure system).
For $D=3$, the system undergoes a paramagnetic-antiferromagnetic
phase transition,
where $m_\text{s}$ is the order parameter.

Let us consider a single sample of the system  (i.e., a fixed 
$\{\epsilon_\bx\}$).
In our tethered computation, we define smooth magnetizations $\hat m$ and $\hat m_\text{s}$
by coupling $m$ and $m_\text{s}$ to Gaussian baths and work in a statistical 
ensemble for fixed $(\hat m,\hat m_\text{s})$ with weight~\cite{fernandez:09}
\begin{equation}\label{eq:weight}
\omega(\hat m,\hat m_\text{s}; \{s_\bx\}) \propto \ee^{-\beta U} \gamma(\hat m,m) \gamma(\hat m_\text{s},m_\text{s}),
\end{equation}
where $\gamma(\hat x, x) = \ee^{N(x-\hat x)} (\hat x-x)^{(N-2)/2}
\varTheta(\hat x-x)$, and $\varTheta(\hat x-x)$ is the step function.
The smoothing procedure shifts the mean value of the parameters, so $\hat x
\simeq x +1/2$.  This ensemble is related to the canonical one through a
Legendre transformation. For instance, the partition function of the system is
\begin{equation}\label{eq:Z}
\begin{split}
Z&= \int\dd \hat m \dd \hat m_\text{s}\  \sum_{\{s_\bx\}} \omega(\hat m,\hat m_\text{s}; \{s_\bx\}) \ \ee^{\beta N (h\hat m + h_\text{s}\hat m_\text{s})} \\
 &=\int \dd \hat m\dd \hat m_\text{s}\ \ee^{-N [\varOmega_N(\hat m,\hat m_\text{s}) -\beta h \hat m - \beta h_\text{s} \hat m_\text{s}]}\, , 
\end{split}
\end{equation}
where $\varOmega_N(\hat m,\hat m_\text{s})$  is the Helmholtz effective potential.

We can reconstruct $\varOmega_N$ from computations at fixed $(\hat m,\hat m_\text{s})$
via the so-called tethered field $(\hat b,\hat b_\text{s})$
\begin{align}\label{eq:hatb}
\hat b &= 1- \frac{1/2-1/N}{\hat m-m}, &
\hat b_\text{s} &= 1- \frac{1/2-1/N}{\hat m_\text{s}-m_\text{s}}.
\end{align}
In particular, the gradient $\boldsymbol\nabla \varOmega_N$ is 
\begin{equation}\label{eq:Omega-hatb}
 \bigl( \partial \varOmega_N / \partial\hat m,\ 
\partial \varOmega_N/\partial \hat m_\text{s} \bigr) = \bigl( \langle \hat b\rangle_{\hat m,
\hat m_\text{s}}, \ \langle \hat b_\text{s}\rangle_{\hat m,\hat m_\text{s}}\bigr).
\end{equation}
The notation $\langle\cdots\rangle_{\hat m,\hat m_\text{s}}$ denotes tethered 
expectation values, computed with weight~\eqref{eq:weight}. 

A TMC computation consists in a set of independent Monte Carlo
simulations at fixed $(\hat m,\hat m_\text{s})$ that are then combined 
to reconstruct $\varOmega_N$. 
Note that the effective potential (as a function of the magnetizations)
has all the information about the system in the tethered
ensemble, just as the free energy (as a function of the applied fields)
has all the information in the canonical ensemble.

The canonical averages at fixed $(h,h_\text{s})$ can be recovered with Eq.~\eqref{eq:Z}.
Note that, according to~\eqref{eq:Omega-hatb}, this integral is dominated by 
\emph{saddle points} $(\hat m,\hat m_\text{s})$ such that
\begin{align}\label{eq:saddle-points}
\langle \hat b\rangle_{\hat m,\hat m_\text{s}} &= \beta h, &
\langle \hat b_\text{s}\rangle_{\hat m,\hat m_\text{s}} &= \beta h_\text{s}.
\end{align}
We can determine the relative weights of different
saddle points by line-integrating the tethered field 
along any connecting path. We are interested in the case $h_\text{s}=0$.

\begin{figure}
\centering
\includegraphics[height=\columnwidth,angle=270]{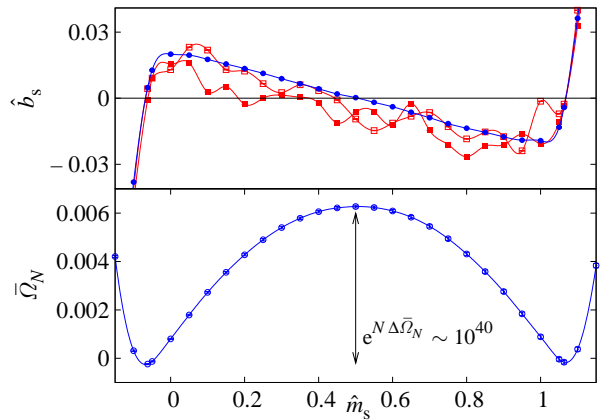}
\caption{(color online) \emph{Top:} Tethered
field $\langle \hat b_\text{s}\rangle_{\hat m,\hat m_\text{s}}$, Eq.~\eqref{eq:hatb},
at $T=1.6$ and $\hat m=0.11$, 
for two individual samples of an $L=24$ system ($\square$ and $\blacksquare$) and
for the sample average ({\Large\textbullet}) as a function of $\hat m_\text{s}$.
The field is  self-averaging in the region outside the two external zeros.
The errors cannot be seen at this scale. 
\emph{Bottom:} Effective potential $\bar\varOmega_N(\hat m=0.11,\hat m_\text{s})$
obtained by integrating the averaged tethered field of the top panel.
The two antiferromagnetic minima are separated by a very large 
barrier (the escape time is $\tau \sim \exp[N\Delta\bar\varOmega]$),
and there is no paramagnetic minimum.}
\label{fig:hatb}
\end{figure}

So far we have summarized the application of TMC for a single sample.
Since it consists of simulations at fixed $(\hat m,\hat m_\text{s})$,
it eliminates the need to tunnel between coexisting phases and, hence, 
equilibrates the system much faster than a canonical simulation.
However, we still face the serious problem of self-averaging violations.
In principle, the definition of quenched disorder implies reconstructing
the free energy with~\eqref{eq:Z} before computing
the disorder average. In this work, however, we 
 sample average the Helmholtz potential rather than the free energy 
(a similar approach was taken  in Ref.~\onlinecite{fernandez:08}).

In order to motivate this approach, let us consider Figure~\ref{fig:hatb}---Top.
We compare the tethered average $\langle \hat b_\text{s}\rangle_{\hat m,\hat m_\text{s}}$
for two individual samples with the disorder average over $1000$ samples. The zeros
of this latter curve separate 
an internal gap with chaotic fluctuations, where the field vanishes
in the thermodynamical limit,
from an external region where the field is actually
self-averaging.

We exploit the situation
by considering a small, but finite, value of $h_\text{s}$.
The saddle point defined by this field will be in the self-averaging region.
We can therefore solve the saddle-point equations~\eqref{eq:saddle-points}
on average, rather than sample by sample. Only afterwards do we make $h_\text{s}\to0$
in the solution (this is analogous to the mathematical
definition of spontaneous symmetry breaking).
The limit $h_\text{s}=0^+$ is essentially equivalent to considering a `smeared'
saddle point and averaging over all $\hat m_\text{s}$
\begin{equation}\label{eq:promedio-hatm}
\overline{\langle O\rangle}_{\hat m} = \int\dd\hat m_\text{s}\ 
\overline{\langle O\rangle}_{\hat m,\hat m_\text{s}}
\ee^{-N [\bar \varOmega_N(\hat m,\hat m_\text{s})-\varOmega_0]}\, .
\end{equation}
$\varOmega_0$ is a normalization constant.
Since we work at fixed $\hat m$, $\bar\varOmega_N$ 
is just the  one-dimensional integral
of $\overline{\langle \hat b_\text{s}\rangle}_{\hat m,\hat m_\text{s}}$.

The other saddle-point equation,
$\overline{\langle \hat b\rangle}_{\hat m}=\beta h$, defines a one-to-one relation $\hat m(h)$
so that $\overline{\langle O\rangle}_{\hat m(h)}$ and the canonical
$\overline{\langle O\rangle}(h)$ both tend to the same thermodynamical limit (ensemble equivalence).
Furthermore, 
for finite lattices  $\overline{\langle O\rangle}_{\hat m}$
is better behaved statistically and 
arguably more faithful to the physics of an experimental sample.
Therefore, we shall identify $\overline{\langle O\rangle}(h)=\overline{\langle O\rangle}_{\hat m(h)}$
and use the more familiar canonical notation. 
See Refs.~\onlinecite{fernandez:09,martin-mayor:11} for a more detailed
study of this ensemble equivalence.

We have used the above outlined procedure to thermalize the DAFF
for temperatures down to $T=1.6$ and sizes up to $L=32$ ($1000$ samples
for $L=8,12,16,24$ and $700$ samples for $L=32$). For each size
we simulate a grid of $\approx150$ points in the $(\hat m,\hat m_\text{s})$
plane ($5$ values of $\hat m$, and $\approx 30$ values of $\hat m_\text{s}$
on each). We also use temperature parallel tempering.
This is only necessary to thermalize $L\geq24$, but it is convenient for smaller 
lattices because we are also interested in the $T$ dependence. Thermalization is ensured
using the methods described in Ref.~\onlinecite{janus:10}.
We provide more technical details in Ref.~\onlinecite{martin-mayor:11}.

The first interesting physical result is the effective potential itself.
Some authors have found metastable behavior
in the DAFF, interpreted as a sign
of a first-order transition.~\cite{sourlas:99,*maiorano:07} This should manifest as the coexistence of 
antiferromagnetic and paramagnetic minima in $\bar\varOmega$.
However, see Figure~\ref{fig:hatb}---Bottom, our results exhibit 
only two antiferromagnetic minima, separated by a very large free-energy barrier.
In a canonical simulation, the system tunnels back and forth between the two,
with an escape time $\tau\sim\exp[N\Delta \bar \varOmega]$. This explains the metastable behavior
observed in previous work (and the difficulty to thermalize large samples
with canonical methods), but is inconsistent with a first-order scenario.

Of course, we could  be looking
at a value of $\hat m$ (equivalently, of $h$) far from the critical point.
In order to find the phase transition, we
compute the usual second-moment correlation length $\xi$.~\cite{ballesteros:96}
We use the propagator  $F_h(\boldsymbol k) = N \overline{\langle \phi(\boldsymbol k) \phi(-\boldsymbol k)\rangle}(h)$,
where  $\phi$ is the staggered Fourier transform
of the spin field.

\begin{figure}
\centering
\includegraphics[height=\columnwidth,angle=270]{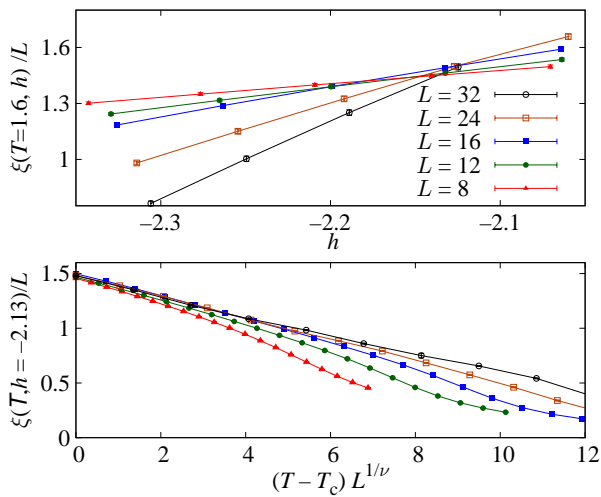}
\caption{(color online) \emph{Top:} Correlation length $\xi/L$ as a function
  of the applied magnetic field $h$ for $T=1.6$. The curves intersect, marking
  a second-order phase transition.  \emph{Bottom:} Scaling plot of $\xi$ as a
  function of $T$ for $h=-2.13$, showing large corrections to leading scaling
  (we use $\nu=1.05$). \label{fig:xi}}
\end{figure}

We have plotted $\xi(h)/L$ at $T=1.6$ as a function of the applied field $h$
in Figure~\ref{fig:xi}---Top.  The curves for different $L$ show very clear
intersections, marking the onset of a second-order phase transition.  In order
to estimate the critical exponents, we apply the quotients
method.~\cite{ballesteros:96}  We consider the ratios of physical observables
for system sizes $(L,2L)$, computed at the intersection point $h^*(L)$ of
their respective $\xi(h)/L$.  We have applied this method to $\partial_h
\xi\sim L^{1+1/\nu_h}$ and $\overline{\langle m_\text{s}^2\rangle }(h)\sim
L^{2\beta/\nu_h-3}$ in Table~\ref{tab:quotients}.  Note that our estimate for
$\beta$ is very low, in accordance with previous numerical and experimental
work.

We can also estimate $\nu$ from the temperature dependence of $\xi$ at fixed $h$,
obtaining a second estimate $\nu_T$ (Table~\ref{tab:quotients}).
Both determinations of $\nu$ should coincide,
but we obtain $\nu_h\approx 0.75$ and $\nu_T\approx1.05$.
We can see in Figure~\ref{fig:xi}---Bottom that 
this discrepancy is due to strong scaling corrections.
If one attempts a
collapse of the curves, focusing on different ranges for $\xi/L$, 
the corresponding values of $\nu$ vary from $\nu\approx0.75$ to $\nu>2$, which
explains the wide range of variation in previous numerical estimates of $\nu$.
By safely locating the critical point and using the quotients method, we have minimized
the scaling corrections, but not eliminated them completely.

\begin{table}
\centering
\begin{ruledtabular}
\begin{tabular}{clllll}
 $L$ & \multicolumn{1}{c}{$h^*(L)$} & 
\multicolumn{1}{c}{$\beta/\nu_h$}&\multicolumn{1}{c}{$\nu_h$} & 
\multicolumn{1}{c}{$\alpha/\nu_h$}& 
\multicolumn{1}{c}{$\nu_T$}   \\
\hline
8 & $-2.178(4)$ & 0.0125(7) & 0.887(5)& 0.0765(25)  & 1.07(9) \\
12& $-2.140(5)$ & 0.0104(5) & 0.790(9)& 0.0781(27)  & 1.01(4) \\
16& $-2.123(3)$ & 0.0119(4) & 0.742(7)& 0.224(4)    & 1.10(15) \\ 
\end{tabular}
\end{ruledtabular}
\caption{Computation of the critical exponents using the quotients method. We extract
our estimates from ratios of physical observables for sizes $(L,2L)$,
computed at the intersection point of $\xi/L$. The first four columns give results
for fixed $T=1.6$ and the last one at fixed $h=-2.13$.}
\label{tab:quotients}
\end{table}

We  need an additional critical exponent
in order fully to characterize  the critical behavior of the DAFF.
This is the hyperscaling violations exponent $\theta$, 
which can be related to the free-energy barrier between the ordered and the disordered
phase: $\Delta F \propto L^\theta$.~\cite{vink:10,*fischer:11}
The computation of these barriers is very difficult with traditional methods, 
but straightforward with TMC. Indeed, we can identify 
$\Delta F$ with the $\Delta \bar\varOmega_N$
between the two saddle points (disordered and antiferromagnetic) 
defined by the critical $h_\text{c}$. 

\begin{table}
\centering
\begin{ruledtabular}
\begin{tabular}{clllc}
 $L$ &\multicolumn{1}{c}{$\Delta F/N$} & Fit range & 
\multicolumn{1}{c}{$\theta$} & $\chi^2/\mathrm{d.o.f.}$\\
\hline
 8  & 0.03382(29) & $L\geq 8\ $ & $1.448(9)$   & 5.56/3 \\
 12 & 0.01756(15) & $L\geq 12$ &  $1.469(13)$      & 0.44/2\\
 16 & 0.01138(9)  & $L\geq 16$ &  $1.461(20) $     & 0.16/1\\
 24 & 0.00608(5)  &  \\
 32 & 0.00392(5)  & \\
\end{tabular}
\end{ruledtabular}
\caption{Computation of the hyperscaling violations exponent $\theta$ from
the free-energy barriers $\Delta F$. We report fits 
to  $\Delta F = AL^{\theta}$, for different ranges, giving the $\chi^2$ 
and the degrees of freedom of each fit. Our preferred final estimate is $\theta=1.469(20)$, 
taking the central value of the fit for $L\geq12$ and the more conservative
error of the fit for $L\geq16$.  \label{tab:theta}}
\end{table}

We can compute this barrier simply by evaluating the line integral of $\bigl(
\overline{\langle\hat b\rangle}_{\hat m,\hat m_\text{s}} - \beta
h_\text{c},\ \overline{\langle\hat b_\text{s}\rangle}_{\hat m,\hat
  m_\text{s}}\bigr)$ along a path joining the two saddle points. We know that
one of them will lie on the line $\hat m_\text{s}=0.5$ ($m_\text{s}\approx0$).
Therefore, we first integrate from the antiferromagnetic saddle point to $\hat
m_\text{s}=0.5$ at fixed $\hat m$. We then integrate at fixed $\hat
m_\text{s}=0.5$ until we reach the disordered saddle point.  We give the
resulting values of $\Delta \bar\varOmega_N = \Delta F/N$ in
Table~\ref{tab:theta}.  Our final estimate is $\theta=1.469(20)$, incompatible
with the $\theta=D-1$ of a first-order phase transition.

Notice that the hyperscaling relation $2-\alpha=\nu(D-\theta)$, 
coupled with our values for $\nu$ and $\theta$, predicts 
not only a divergence of the specific heat, as observed
in experiments, but also
a positive $\alpha$.  We could test this result directly
by computing $C=\partial_h \overline{\langle m\rangle}$.
Unfortunately, the quotients method is ill-suited to 
this quantity, whose scaling is more aptly described as
$C\simeq A + BL^{\alpha/\nu}$.~\cite{ballesteros:98b} Therefore, 
one needs extremely large values of $L$ to reach the 
asymptotic regime $C\sim L^{\alpha/\nu}$. The behavior of the quotients 
in Table~\ref{tab:quotients} is consistent with this expectation. 

It has been proposed that $\theta$ is not independent, but 
given by $\theta = D/2-\beta/\nu$.~\cite{schwartz:85,*schwartz:86}
Combining Tables~\ref{tab:quotients} and~\ref{tab:theta} we
see that our  numerical results  are indeed compatible
with this two-exponent scenario.

We can use our results to comment on the experimental
situation. In an experimental study, the critical exponents are 
computed from fits to the scattering line shape $S(k)=S_\text{d}(k)+S_\text{c}(k)$,
where the two terms distinguish connected and disconnected contributions.
In the two-exponent scenario, strongly supported by our data, 
the most singular term in $S_\text{d}$ is the 
square of $S_\text{c}$. This Ansatz was applied in Ref.~\onlinecite{slanic:99}, yielding
$\nu=0.87(7)$ and $\eta=0.16(6)$. Since $\eta=\theta-1+2\beta/\nu$, however, 
this last value violates hyperscaling bounds and is also
incompatible with our results. Perhaps taking $S_\text{d}=(S_\text{c})^2$
for the whole function, not just its singularity, is an excessive
simplification. Clearly a better theoretical determination of $S(k)$ 
is needed. Our methods are well suited to a direct numerical 
approach to this question.

We have used the tethered formalism to obtain
a comprehensive picture of the critical behavior of the DAFF,
resolving the inconsistencies in previous work.
This method restores self-averaging to the problem and
is capable of handling rugged free-energy landscapes
to equilibrate much larger systems than canonical parallel tempering.
Our simulations show clear signs of a second-order phase transition
and are consistent both with experiments on the DAFF and with analytical
results for the RFIM. The critical exponents $\theta$ and $\beta/\nu$ 
(equivalently, $\eta$ and $\bar\eta$) are computed with a high precision, 
although our simulations were not optimized for the computation of $\nu$
(equivalently, of $\alpha$). We obtain $\nu=0.90(15)$, consistent with a 
positive $\alpha$.

The tethered approach demonstrated in this paper has a very broad scope and we
believe it can be fruitfully applied to many systems featuring large
free-energy barriers. Indeed, it has already been successfully implemented for
hard-spheres crystallization.~\cite{fernandez:11} Other promising avenues are
the study of Goldstone bosons and the equation of state for the $D=3$ spin
glass,~\cite{janus:10b} or equilibrium and aging relaxation in a metastable
phase (e.g., to prevent crystallization of supercooled
liquids, see Ref.~\onlinecite{fernandez:06}).

We thank N.G. Fytas for his comments on our manuscript. Our simulations were performed on the Red Espa\~nola de Supercomputaci\'on
and at BIFI (\emph{Terminus} and \emph{Piregrid}). We acknowledge partial
financial support from MICINN, Spain, (contract no FIS2009-12648-C03) and
from UCM-Banco de Santander (GR32/10-A/910383). DY was supported 
by the FPU program (Spain).

\end{document}